\documentclass[runningheads]{llncs}
  
\usepackage{graphicx}
\usepackage{multirow}
\usepackage{eucal}
\usepackage{multicol}
\usepackage{booktabs}
\usepackage{subfig}
\usepackage{amsmath}
\usepackage{amsmath,amssymb,amsfonts}
\usepackage{algorithm}
\usepackage{textcomp}
\usepackage[table]{xcolor}
\definecolor{lightgray}{gray}{0.9}
\usepackage{booktabs}
\usepackage{lineno}
\usepackage[table]{xcolor}% 
\usepackage{xcolor,soul}
\sethlcolor{lightgray}
\usepackage{gensymb}
\usepackage{amssymb}
\usepackage{color}
\usepackage[colorlinks=true,
            linkcolor=red,
            citecolor=blue]{hyperref}
\usepackage{indentfirst}
\usepackage[noend]{algpseudocode}
\def\algbackskip{\hskip-\ALG@thistlm}
\usepackage{caption}
\usepackage{helvet}
\usepackage{courier}
\usepackage{bm}
\usepackage{epsfig}
\usepackage{enumitem}
\usepackage{rotating}
\graphicspath{{fig/}}
\usepackage[T1]{fontenc}
\usepackage{booktabs}
\usepackage{subfig}
\usepackage{url,longtable,multirow,morefloats,stfloats, floatflt,cancel,tfrupee}
\makeatletter
\AtBeginDocument{\@ifpackageloaded{textcomp}{}{\usepackage{textcomp}}}
\makeatother
\usepackage{colortbl,xcolor}
\usepackage{pifont}
\usepackage[nointegrals]{wasysym}
\usepackage[format=plain,
            labelfont=it,
            textfont=it]{caption}

\newcommand{\repeatthanks}{\textsuperscript{\thefootnote}}

\makeatletter
\def\algbackskip{\hskip-\ALG@thistlm}

\@ifundefined{etal}{}{}

\usepackage{ifxetex}
\ifxetex\else\if@twocolumn\@ifpackageloaded{stfloats}{}{\usepackage{dblfloatfix}}\fi\fi

\graphicspath{{fig/}}
\begin{document}
\title{Self-pruning Graph Neural Network for Predicting Inflammatory Disease Activity in Multiple Sclerosis from Brain MR Images}

% \vspace{-0.3cm}
\author{
    Chinmay~Prabhakar \inst{1} \and
    Hongwei~Bran~Li\inst{1,2,5} \and
    Johannes~ C.~Paetzold \inst{3}\and
    Timo~Loehr \inst{2}\and
    Chen~Niu \inst{4} \and
    Mark~M\"uhlau\inst{5} \and
    Daniel~Rueckert\inst{2,3} \and
  Benedikt~Wiestler\thanks{Contributed equally as senior authors}\inst{5} \and 
  Bjoern~Menze\repeatthanks\inst{1,2} }

% index{Prabhakar, Chinmay} 
% index{Li, Hongwei Bran} 
% index{Paetzold, Johannes C.} 
% index{Loehr, Timo} 
% index{Muehlau, Mark} 
% index{Rueckert, Daniel} 
% index{Wiestler, Benedikt}
% index{Menze, Bjoern}

\institute{Department of Quantitative Biomedicine, University of Zurich, Switzerland \and
 Department of Computer Science, Technical University of Munich, Germany \and
 Department of Computing, Imperial College London \and
 The First Affiliated Hospital of Xi'an Jiaotong University, China \and 
 Klinikum rechts der Isar, Technical University of Munich, Germany \\
\email{Email: chinmay.prabhakar@uzh.ch}}

\maketitle              % 
\begin{abstract}
Multiple Sclerosis (MS) is a severe neurological disease characterized by inflammatory lesions in the central nervous system. Hence, predicting inflammatory disease activity is crucial for disease assessment and treatment. However, MS lesions can occur throughout the brain and vary in shape, size and total count among patients. The high variance in lesion load and locations makes it challenging for machine learning methods to learn a globally effective representation of whole-brain MRI scans to assess and predict disease. Technically it is non-trivial to incorporate essential biomarkers such as lesion load or spatial proximity. Our work represents the first attempt to utilize graph neural networks (GNN) to aggregate these biomarkers for a novel global representation. We propose a two-stage MS inflammatory disease activity prediction approach. First, a 3D segmentation network detects lesions, and a self-supervised algorithm extracts their image features. Second, the detected lesions are used to build a patient graph. The lesions act as nodes in the graph and are initialized with image features extracted in the first stage. Finally, the lesions are connected based on their spatial proximity and the inflammatory disease activity prediction is formulated as a graph classification task. Furthermore, we propose a self-pruning strategy to auto-select the most critical lesions for prediction. Our proposed method outperforms the existing baseline by a large margin (AUCs of 0.67 \emph{vs.} 0.61 and {0.66} \emph{vs.} 0.60 for one-year and two-year inflammatory disease activity, respectively). Finally, our proposed method enjoys inherent explainability by assigning an importance score to each lesion for the overall prediction. Code is available at \url{https://github.com/chinmay5/ms_ida.git}
% Finally, our proposed method enjoys inbuilt explainability by interpreting the selected lesions.
% 
% our proposed method provides inherent explainability by scoring each individual lesion by its importance to the overall prediction. 
% Furthermore, we show that this formulation incorporates inbuilt explainability, enabling us to discover the most significant lesions for the eventual prediction. 
% \keywords{Graph convolutional network. Neuroimaging. Multiple sclerosis. Inflammatory disease activity. Attention.}
\end{abstract}

\section{Introduction}

Multiple Sclerosis (MS) is a severe central nervous system disease with a highly nonlinear disease course where periodic relapses impair the patient's quality of life. Clinical studies show that relapses co-occur with the appearance of new inflammatory MS lesions in MR images~\cite{wattjes20212021,sormani2013mri}, making MR imaging a central element for the clinical management of MS patients. Further, assessing new MS lesions is crucial for disease assessment and therapy monitoring~\cite{filippi2020identifying,hauser2020treatment}. Unfortunately, prevailing therapies often involve highly active immunomodulatory drugs with potentially severe side effects. Hence, it necessitates developing machine learning models capable of predicting the future disease activity of individual patients to select the best therapy.

While recent approaches applied convolutional neural networks (CNN) to directly learn features from MR image space~\cite{durso2022personalized,falet2022estimating,zhang2018multiple,yoo2016deep,tousignant2019prediction}, there remain challenges to obtain an effective global representation to characterize disease status.
We attribute the difficulty of MS inflammatory disease activity prediction to a set of distinct disease characteristics that are well observable in MR images. First, while lesions have a sufficient signal-to-noise profile in images, their variation in shape, size, and number of occurrences amongst patients make it challenging for existing CNN-based methods that process the whole MRI scan in one go.
Second, with advanced age, small areas appearing in the MRI of healthy individuals may resemble MS. As such, it is crucial to not only predict MS but also to identify the lesions deemed consequential for the final prediction.

To solve this problem, we use concepts from geometric deep learning. Specifically, we propose a two-stage pipeline. First, the lesions in the MRI scans are segmented using a state-of-the-art 3D segmentation algorithm~\cite{isensee2021nnu}, and their image features are extracted with a self-supervised method~\cite{prabhakar2023vit}. Second, the extracted lesions are converted into a patient graph. The lesions act as nodes of the graph, while the edge connectivity is determined using the spatial proximity of the lesions. By this representation, we can solve the MS inflammatory disease activity prediction task as a graph-level classification problem.
We argue that formulating the MS inflammatory disease activity prediction in our two-stage pipeline has the following advantages: (1) Graph neural networks can easily handle different numbers of nodes (lesions) and efficiently incorporate their spatial locations. (2) Modern segmentation ~\cite{isensee2021nnu,ronneberger2015u,wang2022mixed} and representation learning methods ~\cite{chen2020simple,prabhakar2023vit,li2021imbalance} are effective tools for lesion detection and allow us to extract pathology-specific features. (3) 
By operating at the lesion level, it is possible to discover the lesions that contribute most to the eventual prediction, making the decisions more interpretable. Thus, our proposed solution can be a viable methodology for MS inflammatory disease activity prediction to handle the associated challenges.

\subsubsection{Contributions.} Our contribution is threefold: (1) we are the first to formulate the MS inflammatory disease activity prediction task as a graph classification problem, thereby bringing a new set of methods to a significant clinical problem. (2) We propose a two-stage pipeline that effectively captures inherent MS variations in MRI scans, thus generating an effective global representation. (3) We develop a self-pruning module, which assigns an importance score to each lesion and reduces the task complexity by prioritizing the critical lesions. Additionally, the assigned per-lesion importance score improves our model's explainability. 

% \vspace{-0.2cm}
\section{Methodology} \label{method}
% \vspace{-0.15cm}
\subsubsection{Overview.}
\begin{figure}[t!]
    \centering
     % trim -> left, lower, right, upper
    \includegraphics[trim=0 35 0 0, clip, width=1\textwidth]{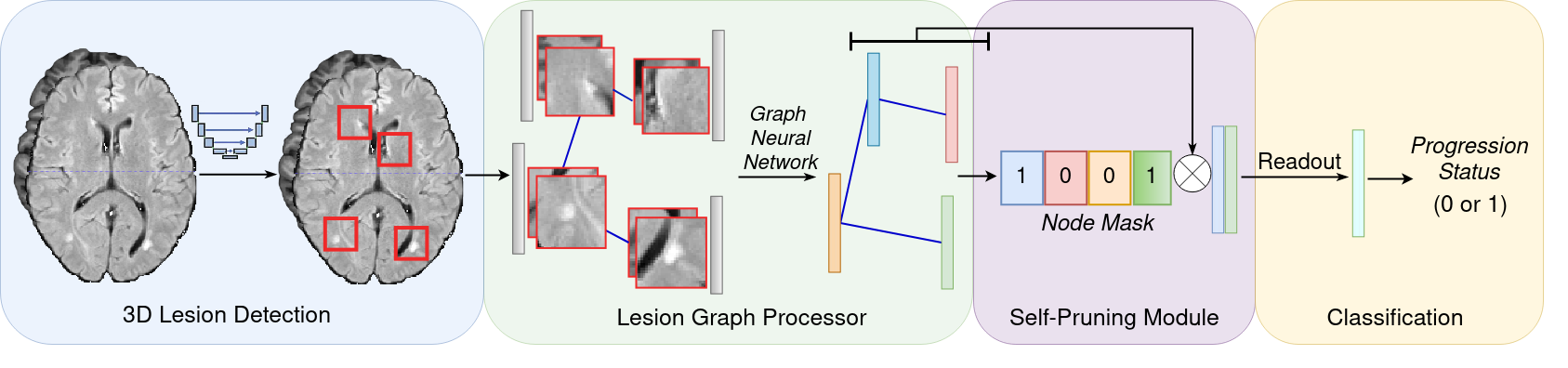}
% \vspace{-0.2cm}
    \caption{\small Our proposed Multiple Sclerosis (MS) inflammatory disease activity prediction framework. We first detect lesions in the MRI scan using nn-Unet~\cite{isensee2021nnu}. A crop centered at the detection is extracted and used to learn self-supervised lesion features. Next, we build a graph from these detected lesions, where each lesion becomes a node, with the connections (edges) between lesions (nodes) defined by spatial proximity. This graph is processed using a graph neural network to generate enriched lesion features. Next, our self-pruning module (SPM) processes these enriched lesion features to determine an importance score for each lesion. The least scoring lesions are pruned-off and the highest scoring lesions are passed to the readout layer to obtain a graph-level feature vector. This graph-level vector is used for the final prediction.}
    \label{fig:main}
    % \vspace{-0.3cm}
\end{figure}

The objective is to predict MS inflammatory disease activity, i.e., to classify if new or significantly enlarged inflammatory lesions appear in the follow-up after the initial MRI scans. 
% considering individual lesions.
We denote the dataset as $D(X, y)$, where $X$ is the set of lesion patches extracted from MR scans, and y $\in$ \{0, 1\} is the inflammatory disease activity status. For patient~$i$, multiple lesion patches \{$x_1^i, x_2^i, ... x_n^i$\} can exist, where $n$ is the total number of lesions. We aim to learn a mapping function $f$: $\{x_1^i, x_2^i, ... x_n^i\}\rightarrow\{0, 1\}$. Please note that our formulation \emph{differs} from existing methods~\cite{tousignant2019prediction,zhang2018multiple,yoo2016deep}, which aim to learn a direct mapping from the MR image to the inflammatory disease activity label. As shown in Fig.~\ref{fig:main}, our proposed method consists of four distinct components. We describe each component in the following sections.
% using the whole MRI scan.

%%%%%%%%%%%%%%%%%%%%%%%%%%
% \vspace{-0.2cm}
%%%%%%%%%%%%%%%%%%%%%%%%%%%
% \vspace{-0.1cm}

% \vspace{-0.3cm}
\subsubsection{Lesion detection and feature extraction.} We focus on the individual lesions instead of processing whole-brain MRI scans. This is important because MS lesions typically comprise less than 1\% of voxels in the MRI scan. With this strategy, the graph model can aggregate lesion-level features for an effective patient-specific representation. %by detecting the lesions and working on the lesion-centered patches.
First, we detect the lesions using a state-of-the-art nn-Unet~\cite{isensee2021nnu} pre-trained with MR images and their consensus annotations from two neuro-radiologists. Then, for each detected lesion, we extract a small fixed-size patch centered at it. Finally, we learn self-supervised features $z$ for the lesion using a recent Transformer-based approach~\cite{prabhakar2023vit}.

% \vspace{-0.3cm}
\subsubsection{Lesion graph processor.} In the second stage, we generate a patient-specific graph $G(V, E, Z)$ from the detected lesions. The lesions act as vertices $V$ of this graph and are initialized with the crop-derived features $Z$. The spatial location $s$ of the lesions is used to determine their connectivity $E$ using a k-Nearest Neighbor (kNN) graph method~\cite{zhang2018multi}. Furthermore, the connected edges are weighted based on their spatial proximity. Specifically, given two lesions $v_i$ and $v_j$, with spatial locations $s_{i}$ and $s_{j}$ respectively, the edge weight $w(v_i, v_j) = exp (- \frac{\|s_{i} - s_{j}\|^2}{\tau^2})$. $\tau$ is a scalar that controls the contribution of distant lesions. Hence the final graph connectivity can be represented as: 

% \vspace{-0.2cm}
\begin{equation} \label{eq:edge_weight}
  E_{i,j}=\left\{
  \begin{array}{@{}ll@{}}
    w(v_i, v_j), & \text{if}\ j \in N(v_i)\ or\ i \in N(v_j)\\
    0, & \text{otherwise}
  \end{array}\right.
\end{equation} 
where $N(v_i)$ are the nodes directly connected to the node $v_i$. Constructing a graph from the lesions is instrumental in two aspects: 
% (1) It enables us to work with MRI data at the patient level where patients have different numbers of lesions. Especially considering that there is no canonical ordering of the lesions; thus, sequential models would struggle. 
(1) The framework allows us to work with varying numbers of lesions in different patients. Alternatively, sequential models could be employed. However, since the lesions lack a canonical ordering, such models would not achieve an effective global representation~\cite{liu2022deep}.
(2) It is possible to incorporate meaningful lesion properties such as spatial proximity and the number of lesions. 

Please note that separate graphs are created for individual patients. Thus, MS inflammatory disease activity prediction is formulated as a graph-level classification task. The graph $G(V, E, Z)$ can be processed using message-passing neural networks (such as GCN~\cite{kipf2016semi}, GAT~\cite{brody2021attentive}, EdgeConv~\cite{wang2019dynamic}, GraphSage~\cite{hamilton2017inductive}, to name a few) to learn enriched lesion features $\hat{Z}$. These enriched features are passed through to the self-pruning module.

% \vspace{-0.3cm}
\subsubsection{Self-pruning module.} The number of lesions can vary substantially among the patients, including the possibility of false positives in the segmentation stage. As such, it is crucial to recognize the most relevant lesions for the final prediction. In addition, this will bring inherent explainability and make it easier for a doctor to validate model predictions. To accomplish this, the enriched lesion features $\hat{Z}$ are passed through a self-pruning module (SPM). The SPM produces a binary mask $M$ for each lesion to determine whether a lesion contributes to the classification. 
% The SPM is realized using a MLP followed by a sigmoid activation. Thus, it generates importance scores for the individual lesions. The high-scoring lesions are retained, while the rest are discarded using
The SPM uses a learnable projection vector $\vec{p}$ to compute importance scores ($\hat{Z} \vec{p}$/${||\vec{p}||}$) for the lesions. These scores are scaled with a sigmoid layer. We retain the high-scoring lesions and discard the rest, which is formulated as:

% \vspace{-0.2cm}
\begin{equation} \label{eq:sip}
\begin{aligned}
% s & =  \\
M &= top{\text -}r(\sigma(\frac{\hat{Z} \vec{p}}{||\vec{p}||}), r) 
\end{aligned}
\end{equation}
Where $\sigma(x) = $ 1/($1 + e^{-x}$) is the sigmoid function and $top{\text -}r(\cdot)$ is an operator which selects a fraction \emph{r} of the lesions based on high importance score. \emph{r} is a hyper-parameter in our setup. Since the masking process is part of the forward pass through the model during both training and inference stages and not a post hoc modification, we refer to it as \emph{self-pruning} of nodes. Features of the remaining nodes ($\hat{Z^\prime} = \hat{Z} \otimes M $) are passed to the classification head.
% through a readout layer. A readout layer aggregates all the node features to produce a single feature vector ($\hat{z^\prime}$) for the entire graph. Finally, these graph-level features are passed through an MLP to obtain the final model prediction $\hat{y}$. We train our model using a binary cross-entropy loss.  

It should be noted that the existence of multiple lesions is a typical characteristic of MS. Hence, a crucial aspect of MS management is that clinicians must identify signs of inflammatory disease activity in MR images to make treatment decisions ~\cite{sormani2013mri}. Therefore, along with predicting inflammatory disease activity, interpreting the contribution of individual lesions to the prediction is essential. By assigning an importance score to each lesion, the SPM can provide explainability to clinicians at a lesion level, while existing CNN methods can not.
% \vspace{-0.3cm}
\subsubsection{Classification head.} The classification head consists of a readout layer aggregating all the remaining node's features to produce a single feature vector $\hat{z^\prime}$ for the entire graph. This graph-level feature is passed through an MLP to obtain the final prediction $\hat{y}$. We train our model using a binary cross-entropy loss.
% \vspace{-0.2cm}
\begin{equation} \label{eq:tot_loss}
\mathcal{L}_{clf} = - \frac{1}{N} \sum_{i=1}^N {(y_i\log(\hat{y_i}) + (1 - y_i)\log(1 - \hat{y_i}))}
\end{equation}
where $N$ is the number of patients, $y_i$ is the ground truth inflammatory disease activity information and $\hat{y_i}$ is the model prediction.

\section{Experiments}
%\subsection{Material}
% \vspace{-0.15cm}
\subsubsection{Datasets and image preprocessing.}
Our approach is evaluated on a cohort of 430 MS patients collected following approval from the local IRB~\cite{hapfelmeier2023retrospective}. Patients included in this analysis were diagnosed with relapsing-remitting MS, with a maximum disease duration of three years at the time of baseline scan. We collect the FLAIR and T1w MR scans for each patient. The scans have a uniform voxel size of 1 x 1 x 1 mm\textsuperscript{3}, were rigidly co-registered to the MNI152 atlas and skull-stripped using HD-BET~\cite{schell2019automated}. Three neuro-radiologists independently read longitudinal subtraction imaging, where FLAIR images from two time points were co-registered and subtracted. In this vein, new and significantly enlarged lesions are identified as positive inflammatory disease activity.

The dataset contains MS inflammatory disease activity information for clinically relevant one-year and two-year intervals~\cite{sormani2013defining}. At the end of the first year, we have the inflammatory disease activity status of 430 patients, with 303 showing activity and 127 not. Similarly, at the end of two years, we have data available for 347 patients, with 287 showing activity and 60 not. Thus, the dataset shows a slight imbalance in favor of inflammatory disease activity. This imbalance is a typical property in the MS patients cohort that impairs algorithm development.

% \vspace{-0.4cm}
\subsubsection{Feature extraction and training configuration.} 
% In this sub-section, we explain each component in the architecture and the training settings in detail.
% \vspace{-0.1cm}
% {\it \flushleft Lesion detection and graph processing.} 
%A random-masking strategy is used to generate masks. The encoder processes only visible, unmasked patches. 
We use an nn-Unet~\cite{isensee2021nnu} for lesion segmentation and detection. Then a uniform crop of size $24 \times24 \times 24 $ mm\textsuperscript{3} is extracted centered at each lesion. The cropped patches are passed through a transformer-based masked autoencoder~\cite{prabhakar2023vit} to extract self-supervised lesion features. %We use a patch size of 2 mm to convert the crop into patches. 
The encoder produces a 768-dimensional feature vector for each patch. We also append normalized lesion coordinates to the encoder output to get the final lesion features.
% which we use as lesion features. We also append normalized lesion coordinates as an additional feature.
% The lesions are connected using a k-nearest neighbor algorithm with $k = 5$. Further, these connections (edges) are weighted using $\tau$ = 0.01 (eq.~\ref{eq:edge_weight}). To enrich the lesion features, the generated graph is processed using two message-passing layers with hidden dimensions 64 and 8, respectively. 
% % These layers have hidden dimensions 64 and 8, respectively.
% Next, the enriched features are passed through the SPM. The SPM uses a learnable projection vector $\vec{p} \in R^8$ and sigmoid activation to learn the importance score. Based on this importance score, a mask is produced to select $r$ = 0.5 (i.e., 50\%) of the highest-scoring lesions and discard the rest. Next, a sum aggregation is used as the readout function. These aggregated features are passed through 2 feed-forward layers with hidden dimensions of size 8. Finally, the features are passed to a sigmoid function to obtain the final prediction.

The lesions are connected using a k-nearest neighbor algorithm with $k = 5$. Further, these connections (edges) are weighted using $\tau$ = 0.01 (eq.~\ref{eq:edge_weight}). Two message-passing layers with hidden dimensions of 64 and 8, respectively, process the generated graph to enrich lesion features. Next, the enriched features are passed through the SPM. The SPM uses a learnable projection vector $\vec{p} \in R^8$ and sigmoid activation to learn the importance score. Based on this importance score, a mask is produced to select $r$ = 0.5 (i.e., 50\%) of the highest-scoring lesions and discard the rest. Next, a sum aggregation is used as the readout function. These aggregated features are passed through 2 feed-forward layers with hidden dimensions of size 8. Finally, the features are passed to a sigmoid function to obtain the final prediction.

The model is trained for 300 epochs using~\emph{AdamW} optimizer~\cite{loshchilov2017decoupled} with 0.0001 weight decay. The base learning rate is 1e-4. The batch size is set to 16. A dropout layer with p=0.5 is used between different feed-forward blocks. Since the dataset is imbalanced in favor of patients experiencing inflammatory disease activity, we use a balanced batch sampler to load approximately the same number of positive and negative samples in each mini-batch.
% \vspace{-0.25cm}
% \subsubsection{Feature extraction.} For each multi-modal volume, we extract a set of 107 traditional radiomics features \cite{van2017computational} including first- and second-order statistics, shape-based features and gray level co-occurrence matrix, denoted as $f_{T}$. For the self-supervised learning one, we extract 768 features from the last transformer head of the encoder ${E}$, denoted as $f_{ViT}$. To directly evaluate the effectiveness of SSL-based features, we concatenate them to a new feature vector $f = [f_{T}, f_{ViT}]$. Note that $f_{T}$ and $f_{ViT}$ are always from the same subjects.

% \vspace{-0.25cm}
\subsubsection{Evaluation strategy, classifier, and metrics.} 
%Our 3D encoder network $E$ is trained in a self-supervised manner and \emph{directly} extract features from raw 3D images without augmentation. 
We report our results on  MS inflammatory disease activity prediction for the clinically relevant one and two-year intervals~\cite{sormani2013defining}. The Area Under the Receiver Operating Characteristic Curve (AUC) is used as the evaluation metric.
% We use \emph{stratified ten-fold nested cross-validation} to reduce selection bias and validate each model. In each fold, we randomly sample 80\% subjects from each class as the training set, 10\% for each class as the validation set, and the remaining 10\% for each class as the test set. The validation set is used to select the best model. 
We use 80\% of the samples as the training set and 10\% as the validation. The validation set is used to select the best model which is then applied to the remaining 10\% cases. This procedure is iterated until all cases have been assigned to a test set once (ten-fold cross-validation). The same folds are used for the proposed model and baseline algorithms.

% %%%%%%%%%%%%%
% \begin{figure}[t]
% 	\begin{center}
% 		\includegraphics[width=0.5\textwidth,height=0.40\textwidth]{fig/main_figure_1_v4.png}
% 	\end{center}
% 	\vspace{-0.2cm}
%     	\caption{AUC score comparison on \emph{BraTS} of the same classification models trained with different kinds of features, including traditional radiomics, self-supervised learning (SSL) based radiomics, the combination of them, and the combination of traditional radiomics with our proposed method. }
% 	\label{fig:AUC_results} 
% \end{figure}
% \vspace{-0.1cm}
% %%%%%%%%%%%%%%%%%%%%%%%%%%%%%%%%%%%%%%%%%%%%%%%%%%%%%%%%%%%%%%%%%
% \vspace{-0.2cm}
\section{Results}
% \vspace{-0.2cm}
\subsubsection{Quantitative comparison.}
Tab.~\ref{tab:results_SOTA} shows the classification performance for the ten folds on one-year and two-year lesion inflammatory disease activity prediction. The $\pm$ indicates the corresponding standard deviations. We compare our method against two existing approaches for MS inflammatory disease activity prediction baselines, a 3D Res-Net~\cite{zhang2018multiple}, and a multi-resolution CNN architecture~\cite{tousignant2019prediction}. These methods learn a direct mapping from the MR image to the inflammatory disease activity label. Our graph model outperforms the baseline methods on one (\textbf{0.67} \emph{vs.} 0.61 AUC) and two-year inflammatory disease activity prediction (\textbf{0.66} \emph{vs.} 0.60 AUC). In the following, we analyze and discuss each component of our established framework.
% \vspace{-0.1cm}

%%%%%%%%%%%%%%%%%%%%%%%%%%%%%%%%%%%%%%%%%%%%%%%%%%
\begin{table*}[t]%[htpb]
% \vspace{-0.2cm}
  \caption[table: Comparison]{\small Comparison of our method against the existing CNN-based solutions. We report the AUC score for MS Inflammatory Disease Activity \emph({IDA)} prediction at the end of one and two years. Our proposed two-stage solution outperforms the existing baselines, achieving the best AUC score on both prediction tasks.}
  \label{tab:results_SOTA}
  \centering
  \setlength{\tabcolsep}{1mm}{
   \begin{tabular}{l | c | c }
    \hline 
 
     &\textit{One year IDA} & \textit{Two year IDA} \\
    {Methods}
    % & 
    % \multicolumn{2}{c|}{Sensitivity} & \multicolumn{2}{c}{Specificity} \\
    % \hline
     ~ & \textit{AUC} & \textit{AUC}           \\
    \hline
     3D ResNet & 0.595 $\pm$ 0.104 & 0.575 $\pm$ 0.099  \\
     CNN multi-res~\cite{tousignant2019prediction} & 0.610 $\pm$ 0.053  & 0.600 $\pm$ 0.059 \\
    \textbf{Ours} & \textbf{0.671} $\pm$ \textbf{0.062} & \textbf{0.664} $\pm$ \textbf{0.063} \\
    
     \hline
  \end{tabular}
  % \vspace{-0.40cm}
}
\end{table*}
%%%%%%%%%%%%%%%%%%%%%%%%%%%%%%%%%%%%%%

\subsubsection{Ablation study.} In this section, we analyze the importance of different components of our proposed method. We defer the analysis of the lesion feature representation to the appendix owing to space constraints.
% \vspace{-0.1cm}
\paragraph{The effectiveness of graph structure.} Since the lesion feature extractor generates rich lesion features, one may argue that the graph structure is unwarranted. There are two alternatives to using a graph, (i) completely discard the graph structure, use a feed-forward layer to enrich the lesion features further, and aggregate them to perform classification~\cite{qi2017pointnet} (Since this formulation regards the input as a set, we call this \emph{Set-Proc} model). (ii) Aggregate all the lesion features for a patient using a mean aggregation and process the aggregated feature by traditional machine learning algorithms such as random forest (RF), support vector machine (SVM) with the RBF-kernel, and logistic regression (LR). Tab.~\ref{tab:trad_methods} compares our model's performance against these alternatives. Our proposed solution obtains better AUC than the alternatives, indicating that incorporating the graph structure is beneficial for eventual prediction.

% \vspace{-0.2cm}
\begin{table}[ht!]
\small
    % \caption{Ablation}
    \begin{minipage}{.48\linewidth}
      \centering
        \captionsetup{width=.95\textwidth}
        \caption{ \small {Effectiveness of the graph structure. We compare our method to traditional ML algorithms and a set-based aggregation baseline, both of which discard the graph structure. The incorporation of the graph structure is beneficial for downstream prediction.}} \label{tab:trad_methods}
        \begin{tabular}{l | c | c }
            \hline 
             &\textit{One year} & \textit{Two year} \\
            {Methods} ~ & \textit{AUC} & \textit{AUC}           \\
            \hline
            SVM & 0.635  \tiny{$\pm$0.05} & 0.513 \tiny{$\pm$0.09} \\
            RF & 0.639  \tiny{$\pm$0.04} & 0.650  \tiny{$\pm$0.09} \\
            LR & 0.658  \tiny{$\pm$0.05} & 0.655  \tiny{$\pm$0.07}\\
            \hline
            Set-Proc & 0.654  \tiny{$\pm$0.06} & 0.658 \tiny{$\pm$0.08} \\
            \hline
            \textbf{Ours} & \textbf{0.671}  \tiny{$\pm$\textbf{0.06}} & \textbf{0.664}  \tiny{$\pm$\textbf{0.06}} \\
            \hline
    \end{tabular}
    
    \end{minipage}%
    \begin{minipage}{.48\linewidth}     
      \centering
      \captionsetup{width=.99\textwidth}
      \caption{\small Importance of spatial proximity. GAT (partially) and TransformerConv (completely) ignore spatial proximity in the graph while GCN, EdgeConv, and GraphSAGE incorporate it. The incorporation of spatial proximity is beneficial for downstream prediction.} \label{tab:spatial}
        \begin{tabular}{l | c | c }
            \hline 
             &\textit{One year} & \textit{Two year} \\
            {Methods} ~ & \textit{AUC} & \textit{AUC}           \\
            \hline
            Tra. Conv~\cite{shi2020masked} & 0.632 \tiny{$\pm$ 0.06}  & 0.627 \tiny{$\pm$ 0.06} \\
            GAT & 0.624  \tiny{$\pm$0.08} & 0.631  \tiny{$\pm$0.09} \\
            \hline   
            Edge & 0.640  \tiny{$\pm$0.09} & 0.657  \tiny{$\pm$0.09} \\
            SAGE & 0.650  \tiny{$\pm$0.05} & 0.634  \tiny{$\pm$0.06} \\
            \textbf{GCN(Ours)} & \textbf{0.671}  \tiny{$\pm$\textbf{0.06}} & \textbf{0.664} \tiny{ $\pm$\textbf{0.06}} \\
            \hline
    \end{tabular}
    \end{minipage} 
\end{table}
%%%%%%%%%%%%%%%%%%%%%%%%%%%

% \vspace{-0.1cm}
\paragraph{The importance of encoding spatial proximity.} The spatial proximity in our model is encoded at two levels. First, lesion connectivity is determined using a NN graph, and second, we weigh the edges based on their distance. Graph convolution layers such as EdgeConv, GCN, and GraphSAGE take the edge weights into account. (EdgeConv does it implicitly by taking a difference of lesion features that already contain spatial information).

On the other hand, the GAT model learns an attention weight and ignores the pre-defined edge weights. However, it still computes these coefficients on only the connected nodes. We can go further, completely ignore the distances and instead use a fully connected graph. The TransformerConv~\cite{shi2020masked} on such a graph is equivalent to applying the well-known transformer encoder ~\cite{vaswani2017attention} on the inputs. Tab.~\ref{tab:spatial} shows that the methods that ignore spatial proximity (TransformerConv) or do not use distance-based weighting (GAT) struggle. On the other hand, EdgeConv, GCN, and GraphSAGE work better. We use GCN in our model owing to its superior performance.
% that both the proposed variations perform worse than methods that incorporate spatial proximity (-vs.-). 

\begin{figure}[t!]
    \centering
    % trim -> left, lower, right, upper
    \includegraphics[trim=25 230 125 150, clip, width=0.99\textwidth]{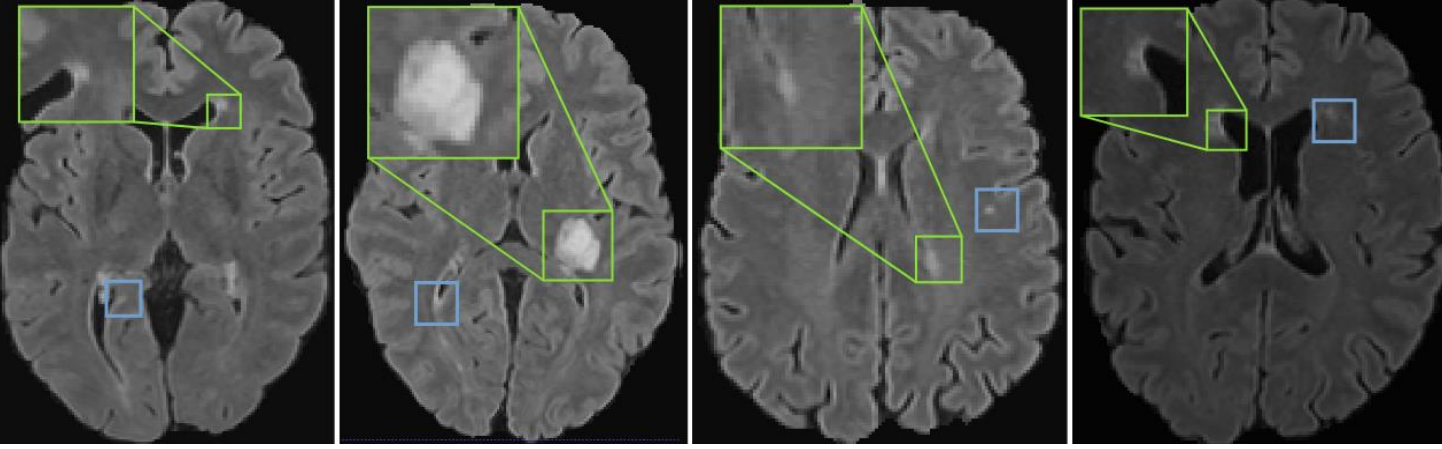}
% \vspace{-0.3cm}
    \caption{\small Lesions selected by the \emph{SPM} for two-year inflammatory disease activity prediction are highlighted with a green bounding box. We also show the zoomed-in view of the lesion. A concurrent lesion in the scan ignored by the \emph{SPM} is shown with a blue bounding box. According to their size and location, the selected lesions are most likely to be relevant to the prediction.}
    \label{fig:viz_sample}
    % \vspace{-0.3cm}
\end{figure}

% \vspace{-0.1cm}
\paragraph{The contribution of the self-pruning module (SPM).} The SPM selects a subset of lesions for the final prediction during the training and evaluation phases. However, the proposed classification method can work without it. In this case, none of the lesions is discarded during the readout operation.
% To verify the importance of the module, we ran an experiment where the SPM was removed for downstream prediction. 
Tab.~\ref{tab:spm_ablation} shows the classification results with and without the SPM. We observe that including SPM leads to better outcomes across different message-passing networks. An explanation could be that the SPM can better handle patient variations (in terms of the total number of lesions) by operating on a subset of lesions (Fig.~\ref{fig:viz_sample}). 
% Additionally, by selecting the most critical lesions, the SPM imparts additional explainability to our model (Fig.~\ref{fig:viz_sample}). 

% \paragraph{Explainability and lesion ranking}
% As discussed in the introduction, a crucial aspect of MS management is that clinicians need to identify the active lesions in MRI images to make treatment decisions \cite{sormani2013mri}. Our approach provides an inherent support for such decisions. In Figure \ref{fig:viz_sample}, we depict an example of the highest ranking lesion, which is evidently a clinically relevant lesion. Essentially, our ranking assigns individual scores to lesions, providing explainability at a lesion level to clinicians, which cannot be provided by whole-scan based CNN classifiers, see baselines.  

%%%%%%%%%%%%%%%%%%%%%%%%%%
% \vspace{-0.2cm}
% \begin{figure}[t!]
%     \centering
%     \includegraphics[width=0.90\textwidth]{fig/examples.png}
% \vspace{-0.3cm}
%     \caption{\small Effect of different loss functions for image reconstruction. We observe the progressive improvement of the reconstructed results from a fraction (25\%) of the input.}
%     \label{fig:reconstruction}
%     \vspace{-0.2cm}
% \end{figure} 
%%%%%%%%%%%%%%%%%%%%%%%%%%%

%%%%%%%%%%%%%%%%%%%%%%%%%%
% \vspace{-0.3cm}
\begin{table}[t]
\small
    % \caption{Ablation}
    \begin{minipage}{.6\linewidth}
      \centering
      \captionsetup{width=.95\textwidth}
      \caption[table: Comparison]{\small Comparison of the performance of different message passing layers with and without the self-pruning module. Incorporating the self-pruning module is beneficial for most message passing layers.}
  \label{tab:spm_ablation}
  \centering
  \resizebox{7.2cm}{!}{
   \begin{tabular}{l | c c | c c }
    \hline 
     &\multicolumn{2}{c|}{\textit{One year}} & \multicolumn{2}{c}{\textit{Two year}} \\
    {Methods}
    ~ & \textit{w/o SPM} & \textit{w/ SPM} & \textit{w/o SPM} & \textit{w/ SPM}        \\
    \hline
    GAT & \textbf{0.628} \tiny{$\pm$\textbf{0.07}} & 0.624 \tiny{$\pm$0.08} & 0.607 \tiny{$\pm$0.06} & \textbf{0.631}\tiny{$\pm$\textbf{0.09}}\\
    Edge & 0.629 \tiny{$\pm$0.07} & \textbf{0.640}  \tiny{$\pm$\textbf{0.09}} & 0.651 \tiny{$\pm$0.07} & \textbf{0.657}  \tiny{$\pm$\textbf{0.09}}\\
     Sage & 0.647 \tiny{$\pm$0.06} & \textbf{0.650} \tiny{$\pm$\textbf{0.05}} & 0.630 \tiny{$\pm$0.07} & \textbf{0.634} \tiny{$\pm$\textbf{0.06}}\\
    \textbf{GCN(Ours)} & 0.640 \tiny{$\pm$0.04} & \textbf{0.671} \tiny{$\pm$\textbf{0.06}} & 0.643 \tiny{$\pm$ 0.07} & \textbf{0.664}\tiny{$\pm$\textbf{0.06}}\\
      
     \hline
  \end{tabular}%
  }
  %\vspace{-0.40cm}
    \end{minipage}%
    \hfill%
    \begin{minipage}{.4\linewidth}     
      \centering
      \captionsetup{width=.99\textwidth}
          \captionof{figure}{\small Model performance against different retention ratio $r$. Best performance observed for $r=0.5$.
          }
            \label{fig:ret}
            % \raggedright 
            % \includegraphics[width=.95\linewidth]{fig/retention_ratio_plot.png}
            % trim -> left, lower, right, upper
            \includegraphics[trim=10 10 10 10, clip, width=4.5cm,height=2.8cm, keepaspectratio]{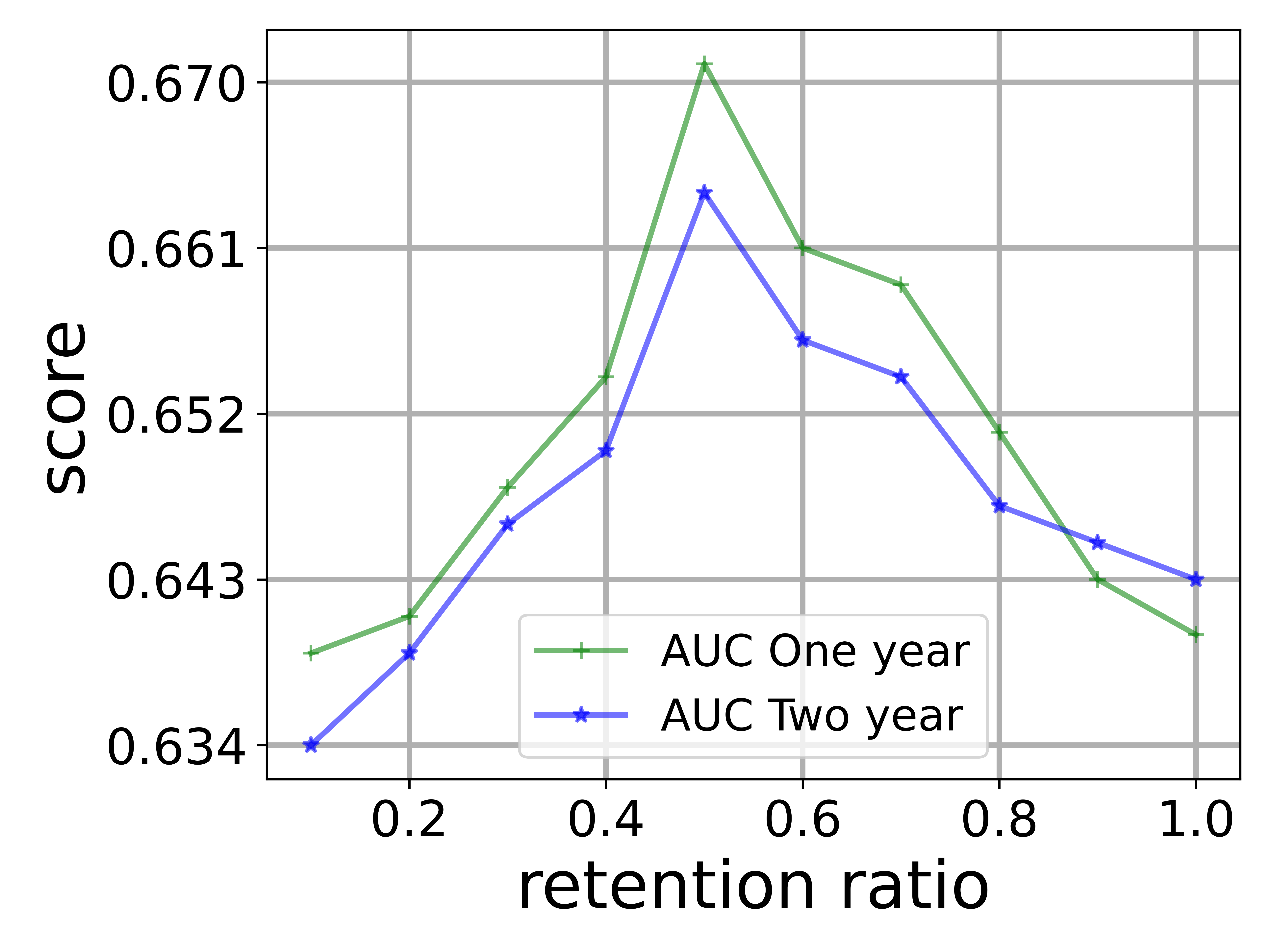}
            % width=10cm,height=10cm,keepaspectratio
            % \vspace{-0.2cm}
    \end{minipage}% 
\end{table}

% \vspace{-0.2cm}
\subsubsection{Analysis of hyperparameters.} The retention ratio \emph{r} and the number of neighbors \emph{k} used for building the graph are the two critical hyperparameters in our proposed framework. We discuss the effect of the retention ratio $r$ here and defer discussion about \emph{k} to the appendix. 
% \vspace{-0.1cm}
{\it \flushleft Effect of retention ratio $r$.} 
The retention ratio $r \in (0, 1]$ controls the fraction of lesions retained after the self-pruning module. If we set its value to 1, all the lesions are retained for the final prediction and thus, bypassing the self-pruning module. Any other value implies that we ignore at least a few lesions in the readout layer. Since the number of lesions can vary across graphs, we retain $\lceil (N. r)  \rceil$ lesions after the self-pruning layer. To find the optimal $r$, we test our model with $r$ between 0.1 and 1.0. The results are summarized in Fig. \ref{fig:ret}. We set $r$ to 0.5 for both tasks.
% This formulation can seamlessly handle cases where the patient has a single lesion (In this scenario, we always retain the lesion).

% \begin{figure}[ht]
%     \centering
%     % trim -> left, lower, right, upper
%     \includegraphics[trim=25 70 15 135, clip, width=0.97\textwidth]{fig/lesion_dist.png}
% \vspace{-0.2cm}
%     \caption{\small Lesion proportion in different brain regions with and without the \emph{SPM} for one-year (left) and two-year (right) progression prediction. We observe that the \emph{SPM} focuses more on the lesions from the juxtacortical region.}
%     \label{fig:viz_sample}
%     \vspace{-0.3cm}
% \end{figure} 

% \vspace{-0.1cm}
\section*{Conclusion}
Predicting MS inflammatory disease activity is a clinically relevant, albeit challenging task. In this work, we propose a two-stage graph-based pipeline that surpasses existing CNN-based methods by decoupling the tasks of detecting and learning rich semantic features for lesions. We also propose a self-pruning module that further improves model generalizability by handling variations in the number of lesions within patients. Most importantly, we frame the MS inflammatory disease activity prediction as a graph classification problem. We hope our work provides a new perspective and leads to cutting-edge research at the intersection of graph processing and MS inflammatory disease activity prediction.
% \vspace{-0.1cm}
% Predicting MS inflammatory disease activity is a clinically relevant, albeit challenging task. In this work, we propose a two-stage graph-based pipeline that achieves superior performance to existing CNN-based methods by decoupling the tasks of detecting and learning rich semantic features for lesions. We also propose a self-pruning module that further improves model generalizability by handling variations in the number of lesions within patients. Most importantly, we frame the MS inflammatory disease activity prediction as a graph classification problem. %We believe this formulation is very beneficial since it opens the door to leveraging progress made in geometric deep learning over recent years. 
\section*{Acknowledgement}
This work was supported by Helmut Horten Foundation. B.W., M.M., and B.M. were supported through the DFG, SPP Radiomics. H.B.L. is supported by an Nvidia GPU research grant. 

\bibliographystyle{splncs04}
%\vspace{-0.4cm}
\bibliography{paper1619}

\begin{thebibliography}{10}
\providecommand{\url}[1]{\texttt{#1}}
\providecommand{\urlprefix}{URL }
\providecommand{\doi}[1]{https://doi.org/#1}

\bibitem{brody2021attentive}
Brody, S., Alon, U., Yahav, E.: How attentive are graph attention networks?
  arXiv preprint arXiv:2105.14491  (2021)

\bibitem{chen2020simple}
Chen, T., Kornblith, S., Norouzi, M., Hinton, G.: A simple framework for
  contrastive learning of visual representations. In: International conference
  on machine learning. pp. 1597--1607. PMLR (2020)

\bibitem{durso2022personalized}
Durso-Finley, J., Falet, J.P., Nichyporuk, B., Douglas, A., Arbel, T.:
  Personalized prediction of future lesion activity and treatment effect in
  multiple sclerosis from baseline mri. In: International Conference on Medical
  Imaging with Deep Learning. pp. 387--406. PMLR (2022)

\bibitem{falet2022estimating}
Falet, J.P.R., Durso-Finley, J., Nichyporuk, B., Schroeter, J., Bovis, F.,
  Sormani, M.P., Precup, D., Arbel, T., Arnold, D.L.: Estimating individual
  treatment effect on disability progression in multiple sclerosis using deep
  learning. Nature Communications  \textbf{13}(1), ~5645 (2022)

\bibitem{filippi2020identifying}
Filippi, M., Preziosa, P., Langdon, D., Lassmann, H., Paul, F., Rovira, {\`A}.,
  Schoonheim, M.M., Solari, A., Stankoff, B., Rocca, M.A.: Identifying
  progression in multiple sclerosis: new perspectives. Annals of neurology
  \textbf{88}(3),  438--452 (2020)

\bibitem{hamilton2017inductive}
Hamilton, W., Ying, Z., Leskovec, J.: Inductive representation learning on
  large graphs. Advances in neural information processing systems  \textbf{30}
  (2017)

\bibitem{hapfelmeier2023retrospective}
Hapfelmeier, A., On, B.I., M{\"u}hlau, M., Kirschke, J.S., Berthele, A.,
  Gasperi, C., Mansmann, U., Wuschek, A., Bussas, M., Boeker, M., et~al.:
  Retrospective cohort study to devise a treatment decision score predicting
  adverse 24-month radiological activity in early multiple sclerosis.
  Therapeutic Advances in Neurological Disorders  \textbf{16},
  17562864231161892 (2023)

\bibitem{hauser2020treatment}
Hauser, S.L., Cree, B.A.: Treatment of multiple sclerosis: a review. The
  American journal of medicine  \textbf{133}(12),  1380--1390 (2020)

\bibitem{isensee2021nnu}
Isensee, F., Jaeger, P.F., Kohl, S.A., Petersen, J., Maier-Hein, K.H.: nnu-net:
  a self-configuring method for deep learning-based biomedical image
  segmentation. Nature methods  \textbf{18}(2),  203--211 (2021)

\bibitem{kipf2016semi}
Kipf, T.N., Welling, M.: Semi-supervised classification with graph
  convolutional networks. arXiv preprint arXiv:1609.02907  (2016)

\bibitem{li2021imbalance}
Li, H., Xue, F.F., Chaitanya, K., Luo, S., Ezhov, I., Wiestler, B., Zhang, J.,
  Menze, B.: Imbalance-aware self-supervised learning for 3d radiomic
  representations. In: Medical Image Computing and Computer Assisted
  Intervention--MICCAI 2021: 24th International Conference, Strasbourg, France,
  September 27--October 1, 2021, Proceedings, Part II 24. pp. 36--46. Springer
  (2021)

\bibitem{liu2022deep}
Liu, C.M., Ta, V.D., Le, N.Q.K., Tadesse, D.A., Shi, C.: Deep neural network
  framework based on word embedding for protein glutarylation sites prediction.
  Life  \textbf{12}(8), ~1213 (2022)

\bibitem{loshchilov2017decoupled}
Loshchilov, I., Hutter, F.: Decoupled weight decay regularization. arXiv
  preprint arXiv:1711.05101  (2017)

\bibitem{nt2019revisiting}
Nt, H., Maehara, T.: Revisiting graph neural networks: All we have is low-pass
  filters. arXiv preprint arXiv:1905.09550  (2019)

\bibitem{prabhakar2023vit}
Prabhakar, C., Li, H.B., Yang, J., Shit, S., Wiestler, B., Menze, B.: Vit-ae++:
  Improving vision transformer autoencoder for self-supervised medical image
  representations. arXiv preprint arXiv:2301.07382  (2023)

\bibitem{qi2017pointnet}
Qi, C.R., Su, H., Mo, K., Guibas, L.J.: Pointnet: Deep learning on point sets
  for 3d classification and segmentation. In: Proceedings of the IEEE
  conference on computer vision and pattern recognition. pp. 652--660 (2017)

\bibitem{ronneberger2015u}
Ronneberger, O., Fischer, P., Brox, T.: U-net: Convolutional networks for
  biomedical image segmentation. In: Medical Image Computing and
  Computer-Assisted Intervention--MICCAI 2015: 18th International Conference,
  Munich, Germany, October 5-9, 2015, Proceedings, Part III 18. pp. 234--241.
  Springer (2015)

\bibitem{schell2019automated}
Schell, M., Tursunova, I., Fabian, I., Bonekamp, D., Neuberger, U., Wick, W.,
  Bendszus, M., Maier-Hein, K., Kickingereder, P., et~al.: Automated brain
  extraction of multi-sequence mri using artificial neural networks. European
  Congress of Radiology-ECR 2019 (2019)

\bibitem{shi2020masked}
Shi, Y., Huang, Z., Feng, S., Zhong, H., Wang, W., Sun, Y.: Masked label
  prediction: Unified message passing model for semi-supervised classification.
  arXiv preprint arXiv:2009.03509  (2020)

\bibitem{sormani2013mri}
Sormani, M.P., Bruzzi, P.: Mri lesions as a surrogate for relapses in multiple
  sclerosis: a meta-analysis of randomised trials. The Lancet Neurology
  \textbf{12}(7),  669--676 (2013)

\bibitem{sormani2013defining}
Sormani, M.P., De~Stefano, N.: Defining and scoring response to ifn-$\beta$ in
  multiple sclerosis. Nature Reviews Neurology  \textbf{9}(9),  504--512 (2013)

\bibitem{thompson2018diagnosis}
Thompson, A.J., Banwell, B.L., Barkhof, F., Carroll, W.M., Coetzee, T., Comi,
  G., Correale, J., Fazekas, F., Filippi, M., Freedman, M.S., et~al.: Diagnosis
  of multiple sclerosis: 2017 revisions of the mcdonald criteria. The Lancet
  Neurology  \textbf{17}(2),  162--173 (2018)

\bibitem{topping2021understanding}
Topping, J., Di~Giovanni, F., Chamberlain, B.P., Dong, X., Bronstein, M.M.:
  Understanding over-squashing and bottlenecks on graphs via curvature. arXiv
  preprint arXiv:2111.14522  (2021)

\bibitem{tousignant2019prediction}
Tousignant, A., Lema{\^\i}tre, P., Precup, D., Arnold, D.L., Arbel, T.:
  Prediction of disease progression in multiple sclerosis patients using deep
  learning analysis of mri data. In: International conference on medical
  imaging with deep learning. pp. 483--492. PMLR (2019)

\bibitem{vaswani2017attention}
Vaswani, A., Shazeer, N., Parmar, N., Uszkoreit, J., Jones, L., Gomez, A.N.,
  Kaiser, {\L}., Polosukhin, I.: Attention is all you need. Advances in neural
  information processing systems  \textbf{30} (2017)

\bibitem{wang2022mixed}
Wang, H., Xie, S., Lin, L., Iwamoto, Y., Han, X.H., Chen, Y.W., Tong, R.: Mixed
  transformer u-net for medical image segmentation. In: ICASSP 2022-2022 IEEE
  International Conference on Acoustics, Speech and Signal Processing (ICASSP).
  pp. 2390--2394. IEEE (2022)

\bibitem{wang2019dynamic}
Wang, Y., Sun, Y., Liu, Z., Sarma, S.E., Bronstein, M.M., Solomon, J.M.:
  Dynamic graph cnn for learning on point clouds. Acm Transactions On Graphics
  (tog)  \textbf{38}(5),  1--12 (2019)

\bibitem{wattjes20212021}
Wattjes, M.P., Ciccarelli, O., Reich, D.S., Banwell, B., de~Stefano, N.,
  Enzinger, C., Fazekas, F., Filippi, M., Frederiksen, J., Gasperini, C.,
  et~al.: 2021 magnims--cmsc--naims consensus recommendations on the use of mri
  in patients with multiple sclerosis. The Lancet Neurology  \textbf{20}(8),
  653--670 (2021)

\bibitem{yoo2016deep}
Yoo, Y., Tang, L.W., Brosch, T., Li, D.K., Metz, L., Traboulsee, A., Tam, R.:
  Deep learning of brain lesion patterns for predicting future disease activity
  in patients with early symptoms of multiple sclerosis. In: Deep Learning and
  Data Labeling for Medical Applications: First International Workshop, LABELS
  2016, and Second International Workshop, DLMIA 2016, Held in Conjunction with
  MICCAI 2016, Athens, Greece, October 21, 2016, Proceedings 1. pp. 86--94.
  Springer (2016)

\bibitem{zhang2018multi}
Zhang, X., He, L., Chen, K., Luo, Y., Zhou, J., Wang, F.: Multi-view graph
  convolutional network and its applications on neuroimage analysis for
  parkinson’s disease. In: AMIA Annual Symposium Proceedings. vol.~2018,
  p.~1147. American Medical Informatics Association (2018)

\bibitem{zhang2018multiple}
Zhang, Y.D., Pan, C., Sun, J., Tang, C.: Multiple sclerosis identification by
  convolutional neural network with dropout and parametric relu. Journal of
  computational science  \textbf{28},  1--10 (2018)

\end{thebibliography}
 \clearpage
\section*{Supplementary Materials}
\begin{table*}[htpb]
% \vspace{-0.2cm}
  \caption[table: Comparison]{\small Comparison of our method against the existing CNN-based solutions. We report the precision, recall and f1 score for the prediction of MS Inflammatory Disease Activity \emph({IDA)} at the end of one and two years. Our proposed two-stage solution outperforms the existing baselines on both prediction tasks.}
  \label{tab:results_SOTA_extend}
  \centering
  \setlength{\tabcolsep}{0.9mm}{
   \begin{tabular}{l |c c c|c c c }
    \hline 
 
     ~&\multicolumn{3}{c|}{\centering\textit{One year IDA}} & \multicolumn{3}{c}{\centering\textit{Two year IDA}} \\
    {Methods} ~ & \textit{prec.} & \textit{recall} & \textit{f1} & \textit{prec.} & \textit{recall} & \textit{f1}  \\
    \hline
     3D ResNet & 0.70 \tiny{$\pm$ 0.03} & 0.75 \tiny{$\pm$ 0.01}  & 0.72 \tiny{$\pm$ 0.06} & 0.73 \tiny{$\pm$ 0.08} & 0.88 \tiny{$\pm$ 0.05} & 0.80 \tiny{$\pm$ 0.03}\\
     
     CNN multi-res & 0.71 \tiny{$\pm$ 0.02}  & 0.78 \tiny{$\pm$ 0.03} & 0.74 \tiny{$\pm$ 0.02} & 0.74 \tiny{$\pm$ 0.02} & 0.90 \tiny{$\pm$ 0.03} & 0.81 \tiny{$\pm$ 0.03}\\
    
    \textbf{Ours} & \textbf{0.78} \tiny{$\pm$ 0.04} & \textbf{0.82} \tiny{$\pm$ 0.03} & \textbf{0.79} \tiny{$\pm$ 0.05} & \textbf{0.85} \tiny{$\pm$ 0.02} & \textbf{0.92} \tiny{$\pm$ 0.08} & \textbf{0.88} \tiny{$\pm$ 0.03}\\
    
     \hline
  \end{tabular}
  % \vspace{-0.40cm}
}
\end{table*}

%\vspace{-1pt}
\begin{figure}[ht]
	\begin{center}
        % trim -> left, bottom, right, top 
		\includegraphics[trim=70 180 150 170, clip, width=0.90\textwidth]{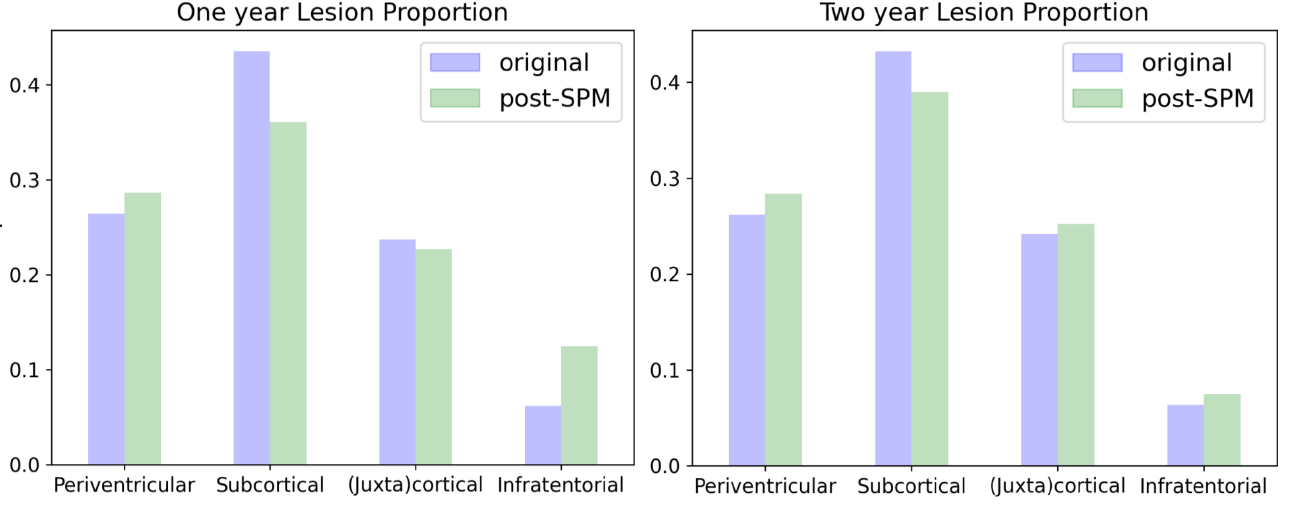}
	\end{center}
	% \vspace{-0.3cm}
    	\caption{MS lesion distribution change across four different brain regions (periventricular, subcortical, (juxta)cortical, and infratentorial) before and after the application of the self-pruning module (SPM) for one-year (left) and two-year (right) inflammatory disease activity prediction. The original distribution (blue) represents the distribution of all the lesions present in the dataset, while post-SPM includes only the lesions considered essential for the eventual prediction. Lesions selected tend to be located in MS-specific regions (periventricular, infratentorial, and (juxta)cortical) and less in subcortical white matter. This behavior is consistent with the McDonald criteria for diagnosing MS~\cite{thompson2018diagnosis}. Hence, our method predominantly selects lesions from clinically relevant brain regions.}
	\label{fig:spm_in_action} 
\end{figure}

% More visualization for the model prediction
\begin{figure}[!h]
	\begin{center}
        % trim -> left, bottom, right, top 
		\includegraphics[trim=0 240 20 100, clip, width=0.98\textwidth]{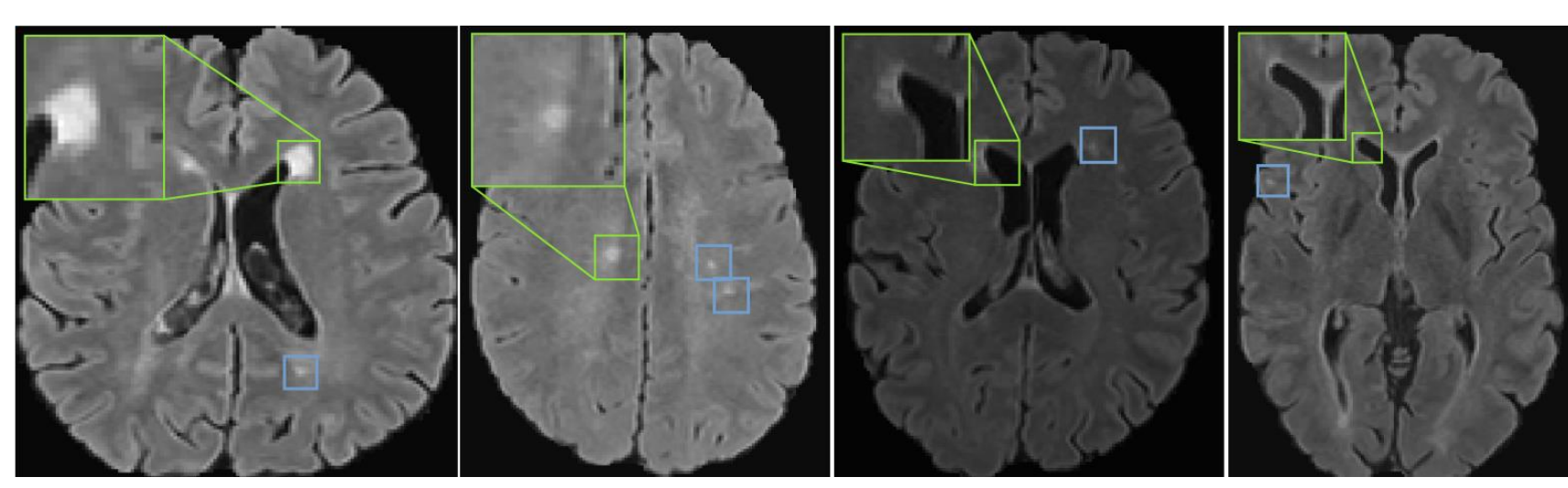}
	\end{center}
	% \vspace{-0.3cm}
         \caption{\small Some more visualizations of the \emph{SPM} in action. Lesions selected by the \emph{SPM} for two-year inflammatory disease activity prediction are highlighted with a green bounding box. We also show the zoomed-in view of the lesion. A concurrent lesion in the scan ignored by the \emph{SPM} is shown with a blue bounding box. According to their size and location, the selected lesions are most likely to be active and clinically relevant to the prediction.}
    	
	\label{fig:viz_appendix}
 \end{figure}

\begin{table*}[htpb]
% \vspace{-0.30cm}
\caption[table: Comparison]{\small Effect of feature representation vs. inflammatory disease activity prediction performance. We report the area under the receiver operating characteristics (AUC) for inflammatory disease activity prediction at the end of one-year and two-year intervals. The cohort consists of T1 and FLAIR MR images for each patient. We run experiments using only the FLAIR, T1, and both modalities to learn self-supervised lesion features. Using lesion features derived from a single modality leads to poor performance of our model, whereas using both modalities to learn these features is advantageous. Hence, enhanced lesion feature representation leads to improved prediction performance.}
  \label{tab:dataset}
  \centering
  \setlength{\tabcolsep}{2mm}{
   \begin{tabular}{l| c| c }
    \hline 
      &\textit{One year progression} & \textit{Two year progression} \\
    Modality & \textit{AUC} & \textit{AUC} \\
    \hline
     FLAIR  & 0.636 \tiny{$\pm$ 0.043} & 0.624 \tiny{$\pm$ 0.066}\\
     T1  &  0.638 \tiny{$\pm$ 0.042} & 0.622 \tiny{$\pm$ 0.067}  \\
     \textbf{Ours(Both)}  & \textbf{0.671} \tiny{$\pm$ \textbf{0.062}} & \textbf{0.664} \tiny{$\pm$ \textbf{0.063}} \\
     \hline
  \end{tabular}
  }
 % \vspace{-0.2cm}
\end{table*}

% \vspace{-0.2cm}
% Figure for k_nn results
\begin{figure}[!ht]
	\begin{center}
        % trim -> left, bottom, right, top 
		% \includegraphics[trim=160 160 260 140, clip, width=0.7\textwidth]{fig/knn_effect.pdf}
        \includegraphics[width=0.8\textwidth]{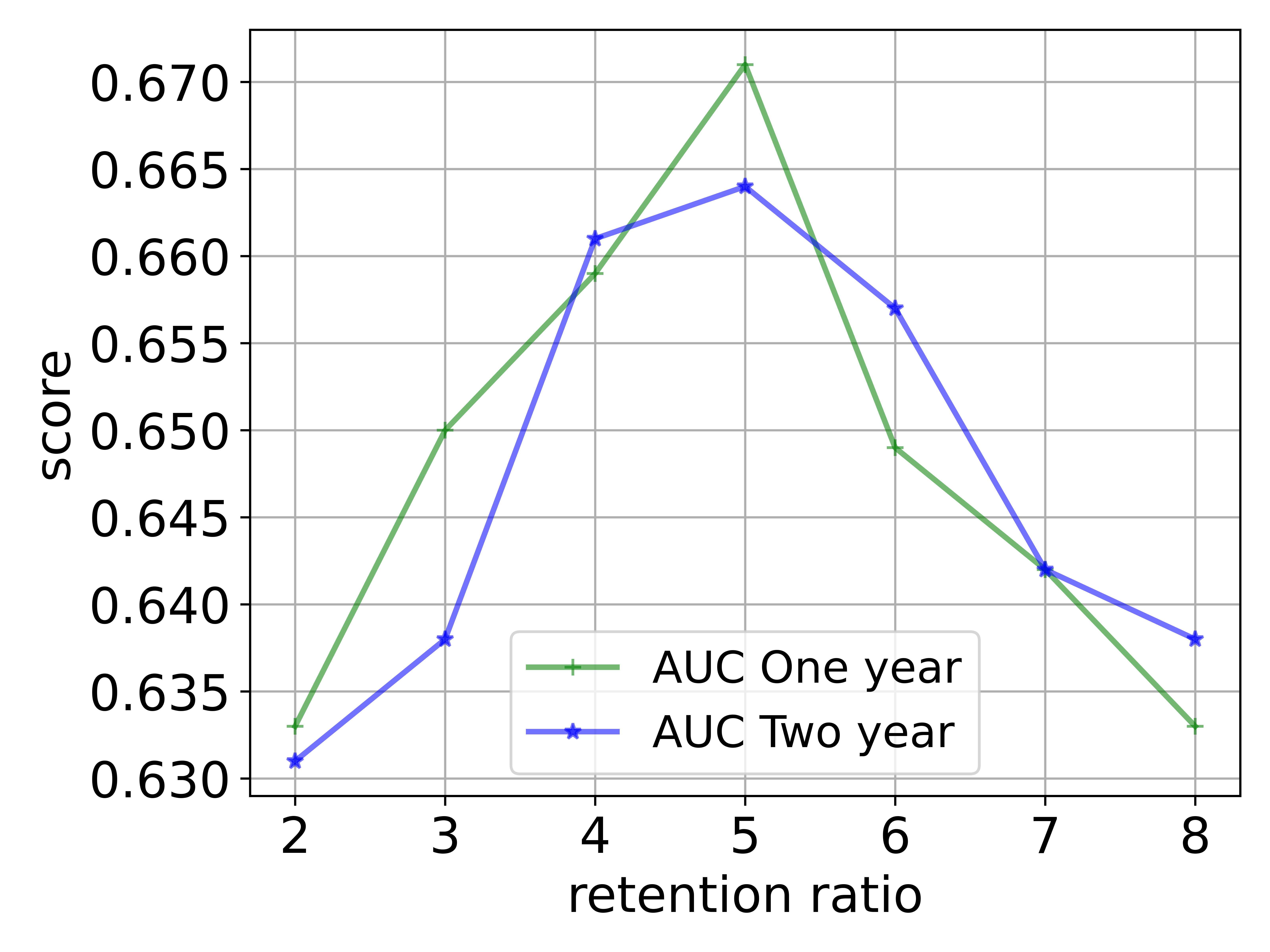}
	\end{center}
	% \vspace{-0.3cm}
    	\caption{The area under the receiver operating characteristics (AUC) \emph{vs.} the number of neighbors \emph{k} used for constructing the k-nearest neighbors (kNN) graph. We show the inflammatory disease activity prediction performance at the end of one and two years. With low values of \emph{k}, the nodes have a small receptive field and, thus, insufficient contextual information. If we increase the value of \emph{k}, our graph slowly degenerates into a fully-connected graph, thus partially losing the spatial context, leading to poor performance. The optimal value of \emph{k} is 5 for predicting inflammatory disease activity at the end of one year and two years.}
	\label{fig:kmeans_results} 
\end{figure}
% \vspace{-0.2cm}

\begin{figure}[h]
	\begin{center}
        % trim -> left, bottom, right, top 
		% \includegraphics[trim=160 140 260 140, clip, width=0.7\textwidth]{fig/num_conv_plot.png}
        \includegraphics[width=0.8\textwidth]{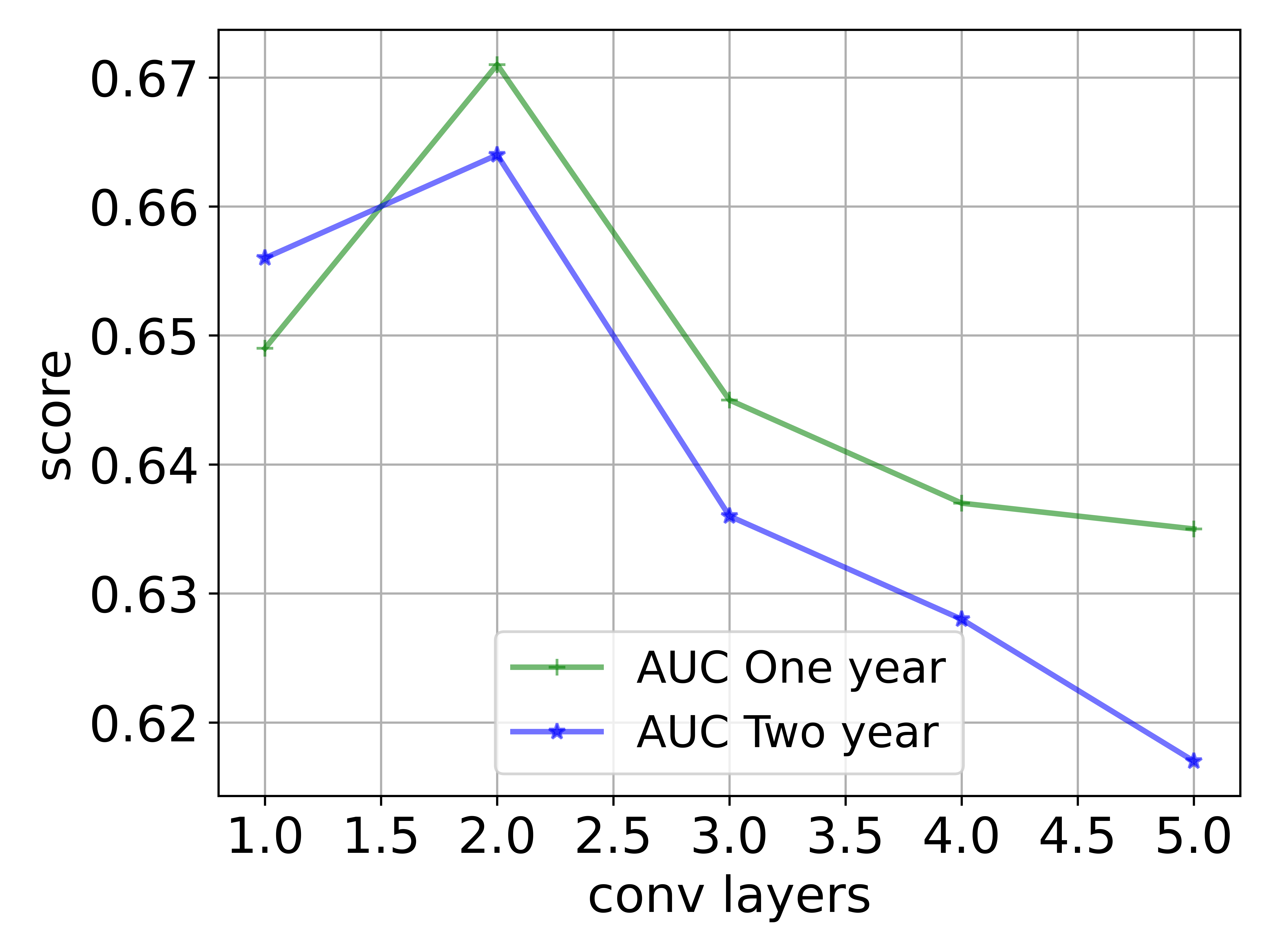}
	\end{center}
	% \vspace{-0.3cm}
    	\caption{The area under the receiver operating characteristics (AUC) \emph{vs.} the number of graph convolution layers (GCN)~\cite{kipf2016semi} used by our model. A single GCN layer aggregates information from the one-hop node neighborhood, and k GCN layers would aggregate information from the k-hop neighborhood. With a single graph convolution layer, the model suffers from the under-reaching effect, i.e., a node's receptive field is limited to only directly connected neighbors, which is insufficient. Hence, we should increase the number of GCN layers to improve model performance. However, if we keep increasing the number of GCN layers, the node features become indistinguishable. This phenomenon is a general drawback of message-passing neural networks and is referred to as the over-smoothing problem~\cite{nt2019revisiting}. Such indistinguishable features lead to deterioration in model performance~\cite{topping2021understanding}. Since we want to balance under-reaching and over-smoothing problems, we use two GCN layers in our final model.}
	\label{fig:num_conv} 
\end{figure}
% \vspace{-0.2cm}

 \clearpage
% \newpage

\end{document}